\documentclass[twocolumn,10pt,aps,prl,preprint]{revtex4}   

\usepackage{amsmath}    
\usepackage{graphicx}   

\begin{document}

\draft
\newcommand{\ajp}{AJP}  

\title{On the spectrum behavior of vibrated granular matter}
\author{Jorge E. Fiscina$^{1}$$^{,}$$^{2}$, Manu\'{e}l O. C\'{a}ceres$^{2}$ and Frank M\"{u}cklich$^{1}$}
\address{$^{1}$Department of Materials Science, Saarland University, D-66123,\\
Saarbr\"{u}cken Germany. \\
$^{2}$Centro At\'{o}mico Bariloche (CNEA), Instituto Balseiro (U.N. Cuyo),\\
and CONICET, 8400, Bariloche, Argentina.}

\date{to be published in the J. of Phys. Cond. Matt.(IOP), in spring 2005}

\begin{abstract}
A laser facility based on a linear image sensor with a sampling period of 100%
$\mu $s allows to investigate the dissipative dynamics of a vibrated
granular matter under gravity. The laser reveals the vertical movement of an
individual Zirconia-Ytria stabilized $2mm$ ball at the surface of a weakly
excited 3D granular matter bed. The stochastic realizations are measured
from the top of the container. Then, power spectra measurements reveal the
different cooperative dynamics of the fluidized gap. We also carried out
measurements for one steel ball and many balls in $1D$ and $3D$ systems. We
fit the measured different regimes with generalized Langevin pictures. We
introduce a fractional temporal operator to characterize the ensemble of
dissipative particles which cannot be represented by a {\it single} Langevin
particle in a complex fluid.

\end{abstract}
\maketitle
\section{Introduction}

Granular media (GM) are ubiquitous in nature and exhibit a wealth of
intriguing physical properties \cite{JNB96}. In this letter we study the
correlations and the statistical properties of vibrated GM under gravity.
Since energy is constantly being added to the system, a non-equilibrium
steady state is reached \cite{BPi99,Blumen94}. We show that in the weakly
excited regime the dynamical behavior of the fluidized particles cannot be
described as {\it simple} Brownian particles, this fact leads us to the
conclusions that the cooperative dissipative dynamics of the GM particles
can be described in terms of generalized Langevin particles.

Hayakawa and Hong \cite{HHo97} introduced the{\it \ }thermodynamics approach
for weakly excited GM. They studied vibrated GM under gravity, and mapped
the non-equilibrium system with a spinless Fermion-like theory. The
important point is that by focusing on the configurational properties of an
excluded volume (hole) theory, the GM non-equilibrium steady state can be
understood in terms of a configurational (maximum principle) Fermi-like
distribution.

Many interesting questions concerning the complex system of GM are still
open, i.e., {\it the} stochastic motion $z(t)$ of the fluidized particles is
not entirely known \cite{GHS92}. For example it is important to test whether
the realization spectrum, $S_{z}(f),$ behaves as a Brownian motion ($1/f^{2}$%
), or it has a more complex behavior related to a cooperative dynamics.
Therefore an analysis of the stochastic realization $z(t)$ of these {\it %
macroscopic }Fermi-like particles ({\it m}Fp) should be made. In this letter
we report some results concerning the analysis of these {\it m}Fp, in
particular we have measured these realizations $z(t)$ and from their Fourier
(FFT) analysis we conclude that $S_{z}(f)\sim f^{-\nu }$ with $\nu =\nu
(f_{e},\Gamma )$, (where $\Gamma =A\omega ^{2}/g$, $A$ is the amplitude and $%
\omega =2\pi f_{e}$ the frequency of the bed oscillations), showing an
agreement with our phenomenological (Langevin) memory-like stochastic
differential equation (SDE) approach.

A sinusoidal vibration is driven by a vibration plate on the GM bed. The
vibration apparatus is set up by an electromagnetic shaker [TIRAVIB$\,5212$%
], which allows for feedback through a piezoelectric accelerometer
(PCB)\cite {Jorge1}. This allow us the control of the frequency
$f_{e}$ and the acceleration in the range of $10$Hz to $7000$Hz,
and $2g$ to $40g$ respectively, where $g$ is the acceleration of
gravity. The control loop is completed by an Oscillator Lab-works
SC$121$ and a TIRA$\,19$/z amplifier of $1$kw.

The $13$ layer GM was set up with $Z_{r}O_{2}-Y_{2}O_{3}$ balls of diameter $%
d=1.99$mm into a glass container of $30$mm of diameter with steel bottom,
see Fig.1(a). Also experiments with steel balls of $d=10.7$mm in different
glass cylinders ($11.2$ and $11.4$ mm) with steel bottom were carried out.

\smallskip In all cases, the stochastic realizations $z(t)$ in the direction
of the acceleration of gravity were measured from the top of cylindrical
containers, during the vibration under gravity and at a given $f_{e}$. The
experiments were carried out in a chamber at $1$atm of air with $5.8\pm 0.2$%
gr/m$^{3}$ of water vapor. The absolute humidity was controlled by using a
peltier condenser and a control loop through a thermo-hygrometer. In the
case of the experiment with $Z_{r}O_{2}-Y_{2}O_{3}$ balls, the humidity is
of major relevance in order to control the contact forces particle-particle
and particle-wall\cite{Jorge2,Knight95}. Under this humidity controlled
conditions, it was not observed surface convection, convection rolls in the
GM bed, nor rotational movement of the bed with respect to the container,
which is typical for a content of water vapor $>10$gr/m$^{3}$.

The realization $z(t)$ of one particle was followed in a 1D window of $12$%
m\smallskip m with a laser device by using a triangulation method,
see Fig.1(a). In Fig.1(b) we show a typical realization $z(t)$
when one steel ball is vibrated at $f_{e}=250$Hz with $\Gamma
=20$;\ under such experimental conditions (sinusoidal excitation
amplitude $A=\Gamma g/\omega ^{2}\sim 79.6\mu $m) it is possible
to observe the quasi non erratic parabolas for the movement of the
ball under gravity. A laser emitter with a spot of $70\mu $m and a
linear image sensor (CCD-like array) enables a high speed
measurement with $100\mu $sec sampling. The linear image sensing
method measures the peak position values for the light spots and
suppress the perturbation of secondary peaks, which makes possible
a resolution of $1\mu $m. The shaker and the laser displacement
sensor were placed on vibration-isolated tables to isolate them
from the external vibrations, and the displacement sensor from the
experiment vibrations. The realization $z(t)$ is a measure of the
oscillations of the distance (difference) between the particle and
the sensor, around the surface of the granular bed (the fluidized
3D gap). The measured realization without excitation reveals a
white noise $<10\mu $m. Then due to the experimental set-up our
effective resolution is no higher than $10\mu $m.
\begin{figure}
  \includegraphics[width=4cm]{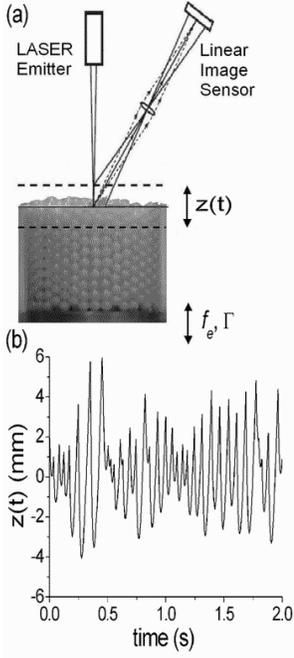}\\
  \caption{(a) Geometry of the distance measurement [realizations of the
amplitude $z(t)$] corresponding to $13$ layer GM with $d=1.99$mm $%
Z_{r}O_{2}-Y_{2}O_{3}$ balls into a glass container of $30$mm of
diameter. (b) A typical realization $z(t)$ for one steel ball
($d=10.7$mm) in a 1D column, when it is performing a quasi non
erratic movement under gravity is shown. In this case
$f_{e}=250$Hz and $\Gamma =20$, thus corresponding to
the relative length $A/d\sim 0.0074$ and dimensionless maximum velocity $%
V_{b}=A\omega /\sqrt{gd}\sim 0.39$. }\label{FIG1:}
\end{figure}
\smallskip The two-second long register of $z(t)$ and the FFT analysis with
Nyquist frequencies of $\ 0.5$ and $1.25$kHz. were taken with a $9354\,$C Le
Croy Oscilloscope of $500$MHz. The velocity $V(t)=\left. dz\right/ dt$ of
the {\it m}Fp and its square dispersion $\sigma _{V}^{2}=\left\langle
V(t)^{2}\right\rangle -\left\langle V(t)\right\rangle ^{2}$ were calculated
from the $z(t)$ registers. In Fig.2(a) we report $\sigma _{V}^{2}$ against $%
\sigma _{z}=\sqrt{\left\langle z(t)^{2}\right\rangle -\left\langle
z(t)\right\rangle ^{2}}$ for fixed $\Gamma =20$ and several $f_{e}$ from $%
200 $Hz to $340$Hz for one steel ball experiments (circles and
squares), and also from $250$Hz to $400$Hz for nine steel balls in
a column. The experiments with many $Z_{r}O_{2}-Y_{2}O_{3}$ balls
($d=1.99$mm) were carried out for fixed $\Gamma =20$ and $f_{e}$
from $70$Hz to $600$Hz, thus
the GM bed was driven by a excitation of amplitude $A$ from $1$mm to $13\mu $%
m. Complementary, one minute films of this fluidized 3D gap were taking by
using a video camera microscope (CVM). For the propose of this work the CVM
was only used to identify qualitatively the different dynamical regimes of
the $Z_{r}O_{2}-Y_{2}O_{3}$ balls, in fact the recorded movies were only
used to discuss the physical meaning of the measured $z(t)$ registers and
their corresponding realization spectra $S_{z}(f)$. These measurements at $%
\Gamma =20$ are a partial research of a large work on progress we carried
out with $5<\Gamma <40$.

\section{The spectrum of the realizations}

When the input of energy is large the individual motion of the beads looks
erratic, in this section we will analyze the frequency spectrum of the
stochastic realizations $z(t)$ of the {\it m}Fp at the fluidized gap. We
remark that in this report we will not be concerned in a regime where the
input of energy is so small that the power spectrum can be described by a
Feigenbaun scenario \cite{Blumen94}.

Consider a given stochastic process (SP) $u(t)$. Then under quite general
conditions if the characteristic function of the SP $u(t)$ (Fourier
transform of the probability distribution $G_{u}(k,t)={\cal F}_{k}\left[
P(u,t)\right] $) obeys the {\it asymptotic} scaling \cite{cac99,libro}:
\begin{equation}
\,G_{u}(\frac{k}{\Lambda ^{H}},\Lambda t)\rightarrow G_{u}(k,t),  \label{s1}
\end{equation}
which means that the SP $u(t)$ fulfills in distribution the scaling $\Lambda
^{-H}\,u(\Lambda t)\rightarrow u(t),$ then, the spectrum density behaves
asymptotically like
\[
S_{u}(f)\propto \left. 1\right/ f^{2H+1}.
\]

Now let us apply this result to a Langevin-like particle. Consider the case
when the velocity $V(t)$ and the position $z(t)$ of the particle are
described by the SDE
\begin{equation}
\frac{dV(t)}{dt}=-\gamma V(t)+\xi (t),\text{\ and}\ \frac{dz}{dt}=V(t),
\label{s3}
\end{equation}
when $\xi (t)$ is a Gaussian colored noise. This SDE corresponds to the
alternative treatment of Brownian motion, in fact initiated by Langevin \cite
{kampen,libro}. This colored noise case corresponds to a weak non-Markovian
process \cite{Cac99b}. From (\ref{s3}) and (\ref{s1}) we obtain, in the
overdamped case (Brownian motion) $S_{z}(f)\propto 1/f^{2}$ (limit $%
f<f_{e}\ll \gamma )$. On the contrary, in the undamped free motion case
(random accelerations model) we obtain $S_{z}(f)\propto 1/f^{4}$ (limit $%
f>f_{e}\gg \gamma $). Nevertheless, in the case when the correlation of the
noise $\xi (t)$ in (\ref{s3}) is of the {\it long-range} type $\left\langle
\xi (t)\xi (0)\right\rangle \propto t^{-\theta }$ with$\ 0<\theta <1$
(perhaps describing a complex fluid surrounding our {\it test} particle),
the SP $z(t)$ turns out to be a {\it strong }non-Markovian process \cite
{Cac99b}, i.e., there is no Fokker-Planck asymptotic regime. Then we get for
the spectrum in the overdamped case
\begin{equation}
S_{z}(f)\propto 1/f^{3-\theta },\text{ if }\left. dz\right/ dt=\xi (t),\text{
with}\ 0<\theta <1,  \label{s7}
\end{equation}
and in the undamped case
\begin{equation}
S_{z}(f)\propto 1/f^{5-\theta },\ \text{if }\left. d^{2}z\right/ dt^{2}=\xi
(t),\text{ with}\ 0<\theta <1.  \label{s8}
\end{equation}

For the particular situation when the ensemble of {\it m}Fp cannot be
represented by a {\it single }particle in a ``fluid'' (Langevin-like
particle) we should leave the regular stochastic calculus and introduce a
{\it fractional} temporal differential operator to emulate the complex
multiple particle collisions and particle motion. Then we model the GM
system by a fractional Langevin equation of the form \cite{BuC04}
\begin{equation}
_{0}D_{t}^{\alpha }\left[ V(t)\right] -V_{0}\frac{t^{-\alpha }}{\Gamma
(1-\alpha )}=-\gamma ^{\alpha }V(t)+\xi (t),\text{ and}\ \frac{dz}{dt}%
=V(t),\ \alpha >0,  \label{s9}
\end{equation}
where $\xi (t)$ is a Gaussian white noise. So, asymptotically in the
undamped limit, $f>f_{e}\gg \gamma ,$ we get
\begin{equation}
G_{V}(\frac{k}{\Lambda ^{H}},\Lambda t)\rightarrow G_{V}(k,t),\text{ with}\
H=\frac{1}{2}-(1-\alpha ),\ \text{when }\alpha \in (\frac{1}{2},1),
\label{s10}
\end{equation}
which means that $S_{V}(f)\propto 1/f^{2\alpha }.$ Then using $\left.
dz\right/ dt=V(t)\ $it follows\
\begin{equation}
S_{z}(f)=\frac{1}{f^{2}}S_{V}(f)\propto 1/f^{2(1+\alpha )},\text{ }\alpha
\in (\frac{1}{2},1),  \label{s11}
\end{equation}
which is the desired result to fit our experiments when friction and
collisions between the GM particles are important issues, and cannot be
represented by a memory-like Brownian model as in Eq.(\ref{s7}), nor by
using the memory-like random acceleration model as in Eq.(\ref{s8}).

The general case given in Eq.(\ref{s9}) by keeping arbitrary the friction
coefficient $\gamma $ and for any noise $\xi (t)$ and correlation $%
\left\langle \xi (t)\xi (0)\right\rangle $ can also be solved in a similar
way \cite{BuC04}.

\section{Results and discussion}
\begin{figure}
  \includegraphics[width=9cm]{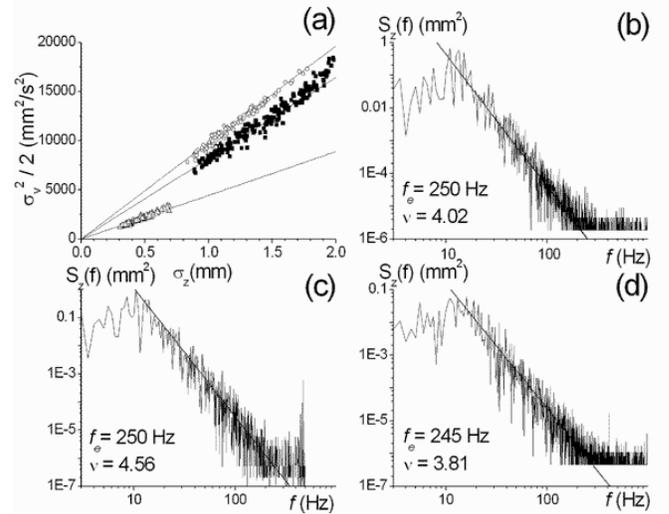}\\
  \caption{(a) $\sigma _{V}^{2}$ as function of $\sigma _{z}$, the circles
correspond to one steel ball, full-squares to one steel ball with
rotation, and triangles to nine steel balls in a $1D$ array.
Typical log-log power spectrum for one steel ball (b-c) and nine
steel balls (d) in a 1D array.}\label{FIG2:}
\end{figure}
In Fig.2(a) we show the velocity squared dispersion $\sigma
_{V}^{2}$ against the amplitude-dispersion $\sigma _{z}$ of the
realizations of the {\it m}Fp corresponding to the $1D$ array of
$10.7$mm steel balls vibrated vertically; for weak amplitude the
slope $\left. \sigma _{V}^{2}\right/ \sigma _{z}$ shows a linear
behavior. Fig.2(a) shows three experiments, the first corresponds
to {\it one steel ball} vertically jumping in a cylinder of
$11.2$mm (almost elastic case) with $S_{z}(f)\sim 1/f^{4},$ see
Fig.2(b), and gives the slope $(9.7\pm 3)m/s^{2}\sim g$ (in accord
with the conservation of energy: $mg\sigma _{z}=\frac{1}{2}m\sigma
_{V}^{2}$). The
second experiment is {\it one steel ball} jumping in a wider cylinder ($11.4$%
mm), in this case the slope gives $(8.2\pm 3)m/s^{2}$ indicating a
dissipative mechanism, here typically $S_{z}(f)\sim 1/f^{4.6}$ see Fig.2(c).
The wider cylinder permits the rotation of the steel ball, probably due to
the interaction with the air gap between the steel ball and the wall. In the
third experiment there are {\it nine steel balls} in the wider cylinder, in
this case the many-body collisions came into account giving a smaller slope $%
(5.9\pm 3)m/s^{2}$, here $S_{z}(f)\sim 1/f^{3.8}$ see Fig.2(d) (we
find that for $400$Hz$>f_{e}>200$Hz, the exponent is $3.5<\nu
<3.8$). This last result indicates that in this regime it is
necessary to introduce a description in terms of a non usual
collision operator. The estimated slope $\left. \sigma
_{V}^{2}\right/ \sigma _{z}$ gives a qualitative magnitude from
which it is possible to observe a clear tendency in the GM motion
behavior. Spectral density measurements [corresponding to the
geometry of Fig.1(a)] associated to the realization $z(t)$ of
$ZrO_{2}-Y_{2}O_{3}$ balls are shown in Fig.3 (a-c). In Fig.3(d)
the plot of the exponent $\nu $ against the excitation frequency
$f_{e}$ for a fixed acceleration $\Gamma =20$ is shown. For
$f_{e}=600$Hz, the balls are organized in a ``lattice'' (we test
this fact by using a VCM at the surface of the GM bed) where the
weak-vibration movement of each ball is limited to a site of this
lattice, given therefore a Wiener realization $z(t)$. Note that
for $\Gamma =20$ and $f_{e}=600$Hz
the excitation amplitude is $A\sim 13.8\mu $m, so we do not expect {\it %
hopping motion} from one to another site. It is interesting to
remark that even when the figures $A/d\sim 0.007$ and $A\omega
/\sqrt{gd}\sim 0.37$ for the Zirconia-Ytria balls are similar to
the one from the experiment with only one steel ball (for $\Gamma
=20$ and $f_{e}=250$Hz), in the present case the stochastic
behavior is quite different as can be corroborated from the value
of $\nu \sim 2$, when it is compared with the one form Fig.2(b-c).
\begin{figure}
  \includegraphics[width=9cm]{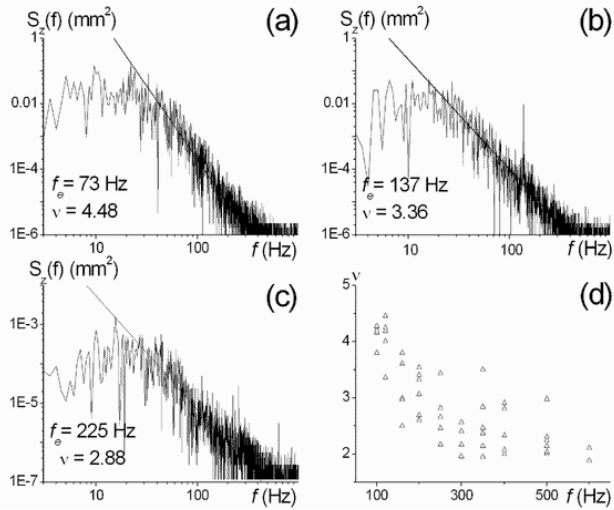}\\
  \caption{Log-log power spectrum for 13 layers of $ZrO_{2}-Y_{2}O_{3}$ balls in
a glass container as in Fig.1(a), for $\Gamma =20$ and different
excitation frequency $f_{e},$ (a) $73$Hz, (b) $137$Hz, (c)
$225$Hz. (d) The power spectrum exponent $\nu $ against the
excitation frequency $f_{e}$ is shown.}\label{FIG3:}
\end{figure}
For $f_{e}$ approximately between $500$Hz and $200$Hz, the
``hopping'' of the fluidized balls (jumping to unoccupied
hole-states) are more frequently, thus leading to exponents $\nu $
from $2$ to $3$. This is a generalized Brownian motion from low to
high correlation, see Eq.(\ref{s7}). When $\nu =3 $ is reached we
observe a pure ``hopping'' mechanism and it is related to a highly
non-Markovian memory in the SDE (limit $\theta \rightarrow 0$),
i.e., the long hopping of one ball is associated with the
long-range noise correlation of the ``fluid''. Such a fluid should
be interpreted as the hopping ``events'' of the whole granular
matter bed. For exponents $\nu $ between $3$ and $4$ (if
$f_{e}\approx 200$Hz to $150$Hz), the motion of a single particle
evolves from a pure hopping to a more complex behavior where the
hopping-balls are perturbed by the collisions from the many-body
bed. This makes necessary the introduction of a new collision
operator which is written in terms of the fractional SDE
(\ref{s9}). In other words, the hopping of a ball is perturbed
(mainly) by few-body collisions. At frequencies $f_{e}$ around
$70$Hz, we observe a pure collision behavior where a ball is under
the random acceleration mechanism transferred from the surrounding
``fluid'', as it is proved because $\nu $ is between $4$ and $5,$
corresponding to the random acceleration dynamics emulated by
Eq.(\ref{s8}).

In general our SDE picture allows us to calculate also the velocity
distribution of the GM, work along this line will be reported elsewhere.

{\it Acknowledgments}. JF thanks A. von Humboldt Foundation and
ADEMAT network. MOC is Senior Associated to the Abdus Salam ICTP.
JF and MOC thank the Director of the ICTP, Prof. K.R. Sreenivasan
for the kind hospitality during our early stay in Trieste, and
Prof. V. Gr\"{u}nfeld for the English revision. FM thanks Krupp
Foundation.

\end{document}